\documentclass[proof]{pasj00}
\draft

\usepackage{times}
\usepackage{textcomp}
\usepackage{threeparttable}

\newcommand{\fermi}{\textit{Fermi}}

\begin{document}

\title{Searching for $\gamma$-Ray Pulsars among \textit{Fermi} Unassociated 
Sources: 2FGL J1906.5+0720}

\author{Yi \textsc{Xing}
and Zhongxiang \textsc{Wang}}
\affil{Shanghai Astronomical Observatory, Chinese Academy of Sciences,\\ 
80 Nandan Road, Shanghai 200030, China}
\email{yixing@shao.ac.cn}
\email{wangzx@shao.ac.cn}


\maketitle

\begin{abstract}

We report the results from our analysis of the \textit{Fermi} Large
Area Telescope data of the \textit{Fermi} unassociated source 
2FGL J1906.5$+$0720, which is a high-ranked candidate pulsar.
In order to better study our target, we first update the ephemeris for PSR J1907$+$0602, which is used to help remove any possible contamination due to strong emission from this nearby pulsar. From our analysis, 2FGL J1906.5$+$0720 is confirmed to have a significant 
low energy cutoff at $\sim$ 1~GeV
in its emission (14$\sigma$--18$\sigma$ significance), 
consistent with those seen in young pulsars.  We search for pulsations 
but no spin frequency signals are found in a frequency range of 0.5--32 Hz.
No single model can fully describe the source's
overall \fermi\ $\gamma$-ray spectrum,
and the reason for this is the excess emission detected at energies of 
$\geq$4 GeV. 
The high-energy component possibly indicates emission from a pulsar 
wind nebula, when considering 2FGL J1906.5$+$0720 as a young pulsar. 
We conclude that 2FGL J1906.5$+$0720 is likely a pulsar based
on the emission properties we have obtained, and observations at other energies are needed
in order to confirm its pulsar nature.

\end{abstract}

\section{Introduction}

Since the \textit{Fermi} Gamma-ray Space Telescope was launched in 
June 2008, the main instrument on-board---the Large Area Telescope (LAT) 
has been continuously scanning the whole sky every three hours in the energy 
range from 20 MeV to 300 GeV, discovering and monitoring $\gamma$-ray 
sources with much improved spatial resolution and sensitivity comparing to 
former $\gamma$-ray telescopes \citep{atw2009}. In 2012 resulting from
\textit{Fermi}/LAT data of the first two-year survey, a catalog of 
1873 $\gamma$-ray sources was released by \citet{nol2012} as 
the \textit{Fermi}/LAT second source catalog. Among the $\gamma$-ray 
sources, approximately 800 and 250 were found to be respectively 
associated with blazars and active galaxies of uncertain types, and more 
than 100 were associated with pulsars in our Galaxy. The three types thus account for the majority of the $\gamma$-ray sources 
detected by \textit{Fermi}. In addition, 575 sources in the catalog have
not been associated with any known astrophysical 
objects \citep{nol2012}. 
For the purpose of identifying the nature of these unassociated sources,
many follow-up studies, such as classifying their $\gamma$-ray 
characteristics \citep{ack2012}, searching for radio pulsars \citep{ray2012},
and observing at multi-wavelengths \citep{tak2012,ace2013}, have been carried 
out.

Because of the relative lack of sources at low Galactic latitudes in many 
extragalactic source catalogs and the emission contamination by
the Galaxy, the Galactic distribution of the \textit{Fermi} unassociated 
sources were found to concentrate towards the Galactic 
plane \citep{nol2012}. More than half of the unassociated sources are 
located at low latitudes with $|b|<$ 10\textdegree \ \citep{nol2012}, 
possibly suggesting Galactic origins for most of them. 
Taking under consideration the types of identified and 
associated Galactic $\gamma$-ray sources in the catalog,
these low-latitude unassociated sources are most likely pulsars, 
pulsar wind nebulae, supernova remnants, globular clusters, or high-mass 
binaries.  Additionally
since identified and associated AGNs or blazars 
have a nearly isotropic distribution, AGN/blazar origins for  
these sources can not be excluded. 
In any case, the low-latitude \textit{Fermi} unassociated sources 
are the best young pulsar candidates on the basis of currently known
Galactic $\gamma$-ray populations, as $\sim$50\% 
of the identified or associated Galactic
\fermi\ sources are pulsars \citep{nol2012} 
and the \textit{Fermi}-detected millisecond pulsars are 
nearly isotropic (see Figure 2 in \cite{lat2013}).
With high rotational energy loss rates (so-called spin-down
luminosities), young pulsars are clustered close to the Galactic plane 
and can be detected to large distances.

Aiming to search for new pulsars among the unassociated sources, we 
selected the pulsar candidates from the \textit{Fermi} second 
source catalog by requiring $|b|<$ 10\textdegree \ and variability indices 
(Variability\_Index parameter in the catalog) lower than 41. 
The variability indices were reported to measure the variability levels 
of sources, and a value larger than 41.64 indicates $<1$\% chance of 
being a steady source \citep{nol2012}. We further ranked the candidates by 
their Signif\_Curve parameters reported in the catalog, which represent 
the significance of the fit improvement between curved spectra and 
power-law spectra, as $\gamma$-ray pulsars typically have curved spectra
with a form of exponentially cutoff power law. The first ten sources from
our selection are listed in Table~\ref{tab:candi}. The first source listed is 
2FGL J1704.9$-$4618, which has the highest Signif\_Curve value 
of $\sim$9.97$\sigma$ but the lowest detection significance value
($\sim$9$\sigma$; Signif\_Avg parameter in the catalog).
For a comparison, the second source in our list 2FGL J1906.5$+$0720 
has both high Signif\_Curve ($\sim$9.85$\sigma$) and Signif\_Avg 
values ($\sim$24$\sigma$), and is ranked the first among candidate
pulsars by \citet{lee2012}, who applied a Gaussian-mixture model 
for the ranking. 
Among the bright $\gamma$-ray sources ($>$20$\sigma$ detection
significance), this source is clearly located in the pulsar region
in the plane of the curvature significance versus variability index
\citep{rom2012}.
We thus carried out detailed study of 2FGL J1906.5$+$0720 by analyzing
\textit{Fermi}/LAT data of the source region, and
report our results in this paper.

In addition, 2FGL J1906.5$+$0720 is located close to a very 
bright $\gamma$-ray pulsar J1907$+$0602 (Signif\_Avg $\sim$ 55$\sigma$;
\cite{lat2013}). The angular distance between them is approximately 
1.3 degrees (see Figure~\ref{fig:ts}). The pulsar was discovered 
in the first $\sim$4 month 
LAT data, revealed with a spin frequency of $\sim$9.378 Hz and a 
spin-down luminosity of 
$\sim$2.8$\times$10$^{36}$ erg s$^{-1}$ \citep{abd2009}. 
The pulsar is radio faint, making very difficult to study its timing behavior 
at radio frequencies \citep{abd2010}.
In order to better study our targeted \fermi\ source by removing 
possible contamination from PSR J1907$+$0602, we performed 
timing analysis to the LAT data of the pulsar and include our timing results in this paper.

\section{Observations}
\label{sec:obs}

LAT is the main instrument on-board the \textit{Fermi} Gamma-ray Space 
Telescope. It is a $\gamma$-ray imaging instrument which carries out an all-sky 
survey in the energy range from 20 MeV to 300 GeV \citep{atw2009}. 
In our analysis we selected LAT events inside 
a 20\textdegree $\times$ 20\textdegree \ region centered at the position of 
2FGL J1906.5$+$0720 during a nearly five-year time period 
from 2008-08-04 15:43:36 to 2013-07-23 20:53:17 (UTC) 
from the \textit{Fermi} Pass 7 database. Following recommendations of 
the LAT team, events included were required to have event zenith
angles fewer than 100 deg, preventing contamination from the 
Earth's limb, and to be during good time intervals when 
the quality of the data was not affected by the spacecraft 
events.

\section{Analysis and Results}
\label{sec:ana}

\subsection{Timing Analysis of PSR J1907$+$0602}
\label{subsec:timing}

After the \fermi\ discovery of PSR J1907$+$0602 \citep{abd2009}, 
its timing solution was updated by \citet{abd2010}
and \citet{ray2011} using the LAT data during 
MJD 54647--55074 and MJD 54682--55211, respectively. 
In 2013 the \textit{Fermi}/LAT team released the second \textit{Fermi} 
catalog of $\gamma$-ray pulsars \citep{lat2013}, in which the timing 
solution for PSR J1907$+$0602 was updated again using the 
data during MJD 54691--55817. A glitch at MJD 55422 was detected 
with $\Delta\nu/\nu$ of $\sim$ 4.6 $\times$ 10$^{-6}$ 
and $\Delta\dot{\nu}/\dot{\nu}$ of $\sim$ 1 $\times$ 10$^{-2}$. 

In order to study 2FGL J1906.5$+$0720 by
being able to remove photons from the nearby pulsar, we performed 
phase-connected timing analysis to the LAT data of J1907$+$0602 during
the nearly five-year time period of MJD 54683--56497. 
We selected LAT events within 0\fdg7 centered at the pulsar's position
given in the catalog in the energy range from 50 MeV to 300 GeV, 
which was suggested by \citet{ray2011}. Pulse phases 
for photons before MJD 55400 were assigned according to 
the known ephemeris using the \textit{Fermi} plugin of 
TEMPO2 \citep{edw2006,hob2006}. We extracted 
an `empirical Fourier' template profile,
with which we generated the time-of-arrivals (TOAs) 
of 128 evenly divided observations of the time period. 
Both the template and TOAs were obtained 
using the maximum likelihood method described in \citet{ray2011}. 
From the pre-fit residuals we found that the timing model given in the second \textit{Fermi} 
catalog of $\gamma$-ray pulsars 
could not fully describe the TOAs after MJD $\sim$55800, suggesting the requirement of an updated timing model. 
We then iteratively fitted the TOAs to the timing model using TEMPO2.
For the glitch because of its relative large amplitude and long interval between the last pre-glitch and the first post-glitch observations, we could not obtain a unique solution to accurately determine its epoch by requiring continuous pulse phase. Instead, we adopted one of the solutions according to the ephemeris we obtained as the glitch epoch, which is closest to that reported in the second \textit{Fermi} catalog of $\gamma$-ray pulsars. The updated ephemeris is given in Table~\ref{tab:timing}, the post-fit timing residuals are shown in Figure~\ref{fig:rms}, and the folded pulse profile and the two-dimensional 
phaseogram of this pulsar are plotted in Figure~\ref{fig:ftp}.

We defined phase 0.1--0.7 as the onpulse phase interval and
phase 0.7--1.1 as the offpulse phase interval (Figure~\ref{fig:ftp}),
using the definition given in \citet{lat2013}.

\subsection{Maximum Likelihood Analysis}
\label{subsec:likeli}

\subsubsection{Full data}

We selected LAT events in an energy range from 100 MeV to 300 GeV 
for the likelihood analysis, and included all sources within 15 degrees 
centered at the position of 2FGL J1906.5$+$0720 in 
the \textit{Fermi} 2-year catalog to make the source model. The spectral 
function forms of the sources are given in the catalog. 
The spectral normalization parameters for the sources within 4 degrees 
from 2FGL J1906.5$+$0720 were left free, and all the other parameters 
were fixed to their catalog values. In addition we included 
the spectrum model gal\_2yearp7v6\_v0.fits and the spectrum 
file iso\_p7v6source.txt in the source model to consider 
the Galactic and the extragalactic diffuse emission, respectively. 
The normalizations of the diffuse components were left free.

In the \textit{Fermi} 2-year catalog, the $\gamma$-ray emission 
from 2FGL J1906.5$+$0720 is modeled by a log parabola expressed by 
$dN/dE=N_{0}(E/E_{b})^{-(\alpha+\beta \ln(E/E_{b}))}$ \citep{nol2012}. 
We fixed the break energy to the catalog value of $\sim$1 GeV, 
and let the indices $\alpha$ and $\beta$ free. We also tested two other
models for the source: an exponentially cutoff power law expressed by 
$dN/dE=N_{0}E^{-\Gamma}exp[-(E/E_{cut})^{b}]$, where $\Gamma$ is 
the spectral index, $E_{cut}$ is the cutoff energy, and $b$ represents the sharpness of the cutoff, and a simple 
power law expressed by $dN/dE=N_{0}E^{-\Gamma}$. 
For the exponentially cutoff power law we note that all pulsars in the 
second \fermi\ 
$\gamma$-ray pulsar catalog with $b$ values different from 1 (usually smaller than 1 and indicating a sub-exponential cutoff) have $E_{cut}$ higher than 2 GeV \citep{lat2013}. Considering the $E_{cut}$ values we obtained for this source are lower than 2~GeV, especially when the possible contamination from nearby sources is excluded (see Section 3.2.2, Section 3.4, and Table~\ref{tab:likeli}), we only used the simple exponentially cutoff shape with $b= 1$ in our analysis.
We performed standard binned likelihood analysis with the LAT science tool 
software package v9r31p1. The obtained spectral results and Test 
Statistic (TS) values are given in Table~\ref{tab:likeli}, and the TS map 
of a $\mathrm{5^{o}\times5^{o}}$ region around 
2FGL J1906.5$+$0720 is displayed in the left panel of Figure~\ref{fig:ts}. 
PSR J1907$+$0602 is kept
in the figure to show the proximity of the two sources.

From the analysis, we found that the log parabola and 
the exponentially cutoff power law better fit the LAT data of 
2FGL J1906.5$+$0720 than the simple power law, indicating 
a significant cutoff in the $\gamma$-ray spectrum of 2FGL J1906.5$+$0720 
at the low energy of $\sim$1 GeV (Table~\ref{tab:likeli}).
The significance of the break (approximately described by $\sqrt{TS_{break}}\sigma$ $=$ $\sqrt{TS_{LP} - TS_{PL}}\sigma$) 
of the log parabola is $\sim$16$\sigma$,  and the significance of 
the cutoff (approximately described by 
$\sqrt{TS_{cutoff}}\sigma$ $=$ $\sqrt{TS_{PL+cutoff} - TS_{PL}}\sigma$) 
of the exponentially cutoff power law is $\sim$14$\sigma$.  

\subsubsection{Offpulse phase intervals of PSR J1907$+$0602}

Considering no offpulse $\gamma$-ray emission from PSR J1907$+$0602 was detected by \textit{Fermi} \citep{ack2011},  we repeated binned likelihood analysis described above by
including LAT events only during the offpulse phase intervals
to prevent possible contamination from the pulsar. The phase intervals are
defined in Section~\ref{subsec:timing}.
Since the emission from the pulsar was 
removed, we excluded this source from the source model. 
The likelihood fitting results 
for the different $\gamma$-ray spectral models for 2FGL J1906.5$+$0720 
are given in Table~\ref{tab:likeli}, and the TS map of a 
$\mathrm{5^{o}\times5^{o}}$ region around 2FGL J1906.5$+$0720 is shown 
in the right panel of Figure~\ref{fig:ts}. The TS values 
are significantly increased comparing to those when the full data were used,
having doubled the detection significance of 2FGL J1906.5$+$0720.
In addition, a low-energy break or cutoff at $\sim$1 GeV in the source's
emission is similarly favored as that in the analysis of the full data.

\subsection{Spectral Analysis}
\label{subsec:sa}

To obtain a spectrum for 2FGL J1906.5$+$0720, we
evenly divided 20 energy ranges in logarithm from 100 MeV to 300 GeV,
and used a simple power law to model the emission in each divided energy range. 
The index of the power law was fixed to the value we obtained before (Table~\ref{tab:likeli}). 
This method is less model-dependent and provides a good description
for the $\gamma$-ray emission of a source. The spectra
from both the full data and the offpulse phase interval data were
obtained, which are displayed in Figure~\ref{fig:spec}. 
Only spectral points with TS greater than 4 (corresponding to the detection significance of 2$\sigma$) were kept.

We plotted the obtained exponentially cutoff power-law fits and
log-parabolic fits from the above likelihood analysis in 
Figure~\ref{fig:spec}.
As can be seen, the first model does not provide a good fit to the LAT spectrum.
At energies of greater than several GeV, the fit deviates from the spectrum
for both the full data and the offpulse phase interval data of PSR J1907$+$0602. 
The log parabola better describes the spectra, 
which is also indicated by the larger TS values obtained with it 
(Table~\ref{tab:likeli}), although a small degree of deviations from
the spectra can still be seen. 
These may suggest an additional spectral component at the high energy range. 

We fit the spectral data points below 2 GeV with exponentially cutoff 
power laws and obtained $\Gamma$ of 1.4$\pm$0.2 and $E_{cutoff}$ of 
1.0$\pm$0.2 GeV from the full data, and $\Gamma$ 
of 1.6$\pm$0.1 and $E_{cutoff}$ of 0.7$\pm$0.1 GeV from the offpulse 
phase interval data. The cutoff energy values are
within the range of young $\gamma$-ray pulsars (0.4 $< E_{cutoff} <$ 5.9;
see Table~9 in \cite{lat2013}) but lower than  
that of millisecond $\gamma$-ray pulsars (1.1 $< E_{cutoff} <$ 5.4;
see Table~9 in \cite{lat2013}).
The fitting again shows that an additional spectral component is 
needed.

\subsection{Spatial Distribution Analysis}

\begin{figure}
\begin{center}
\includegraphics[width=3in]{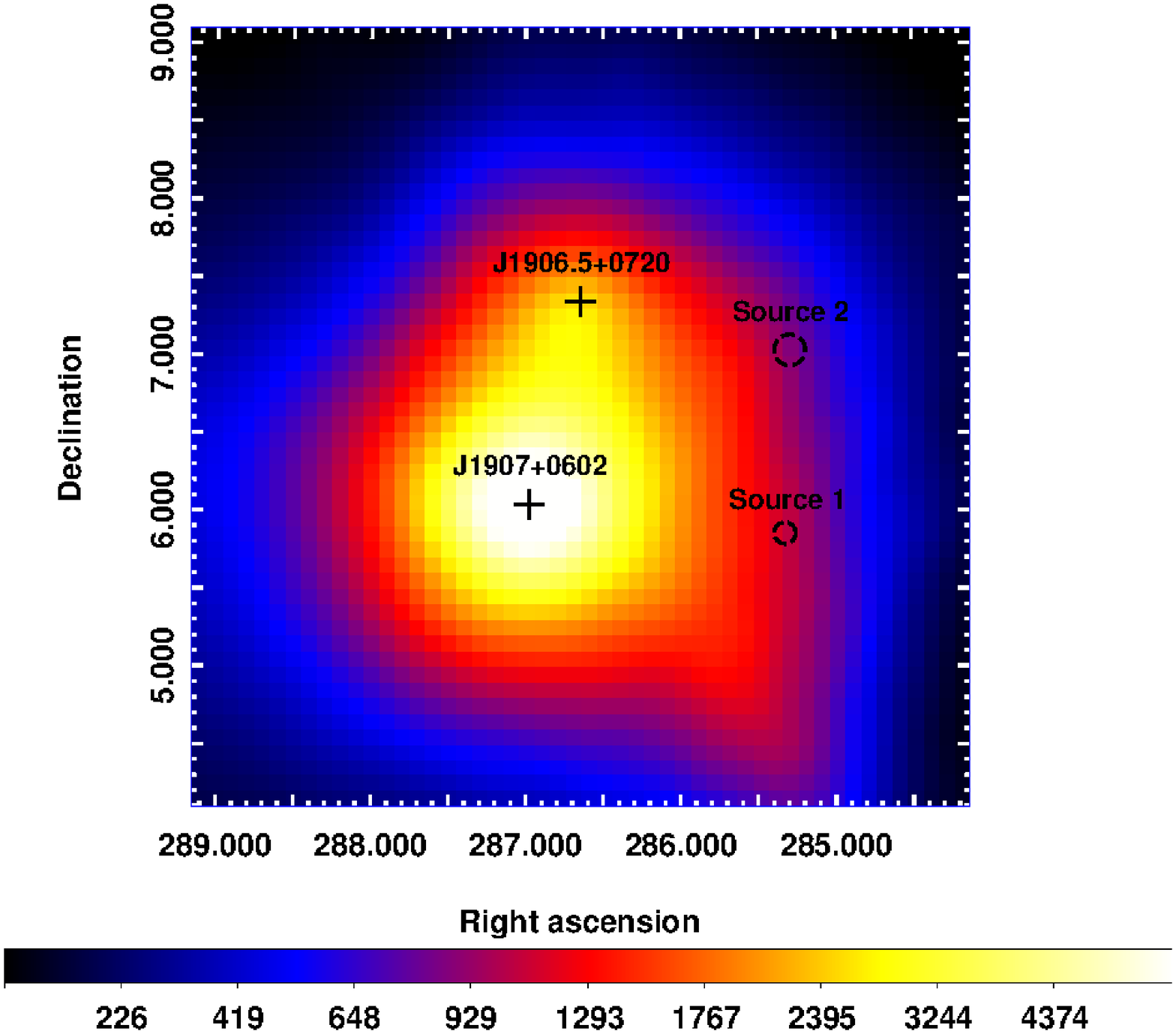}
\includegraphics[width=3in]{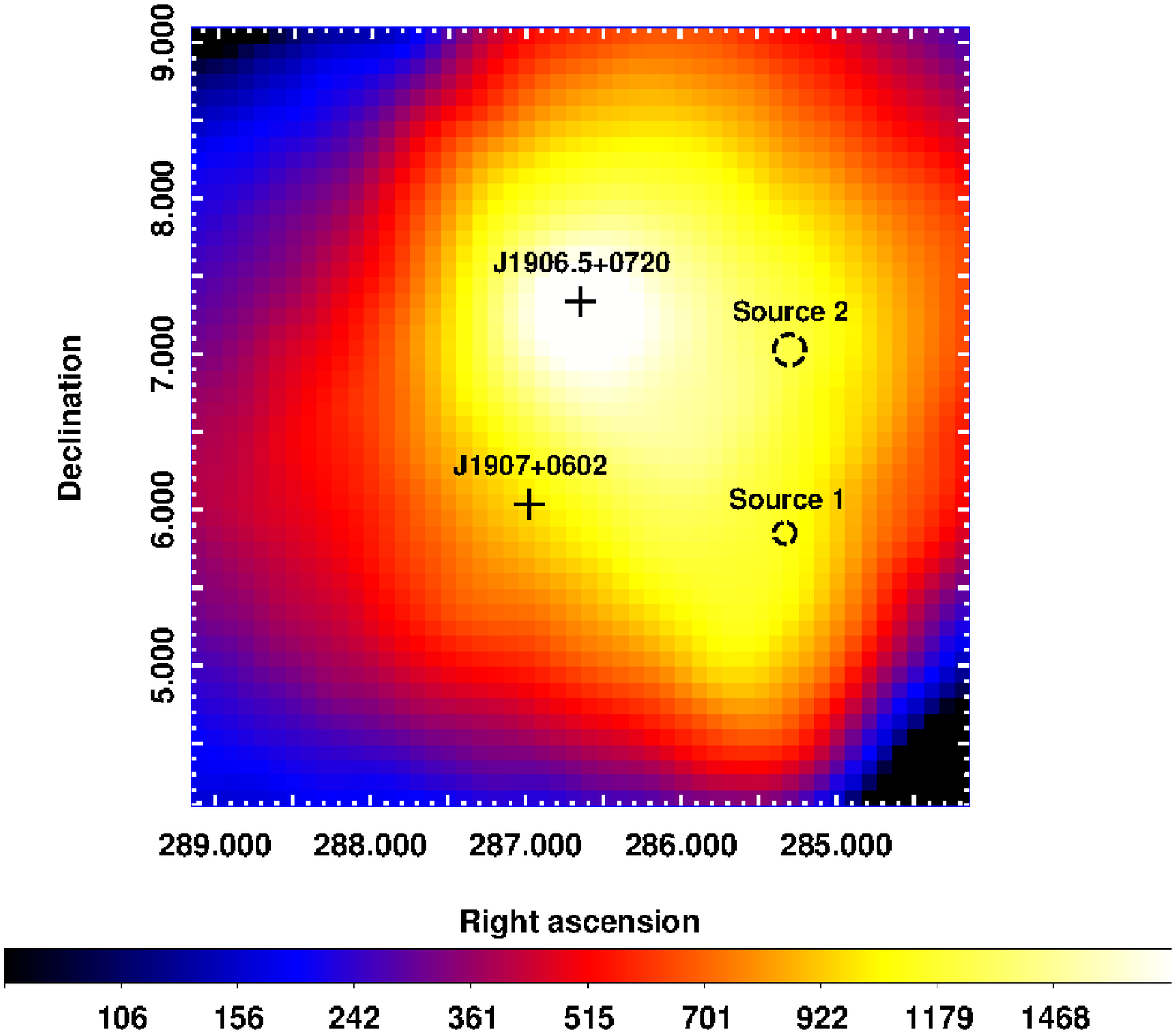}
\end{center}
\caption{TS maps (0.1--300 GeV) of $\mathrm{5^{o}\times5^{o}}$ regions 
centered at R.A.= 286.647$\mathrm{^{o}}$, Decl.= 6.6$\mathrm{^{o}}$ 
(equinox J2000.0) extracted from the full data ({\it left} panel) 
and offpulse phase interval data ({\it right} panel) of PSR J1907$+$0602. 
The image scales of the maps are 0\fdg1 pixel$^{-1}$. Two putative
nearby sources are marked by circles.}
\label{fig:ts}
\end{figure}

\begin{figure}
\begin{center}
\includegraphics[width=5in]{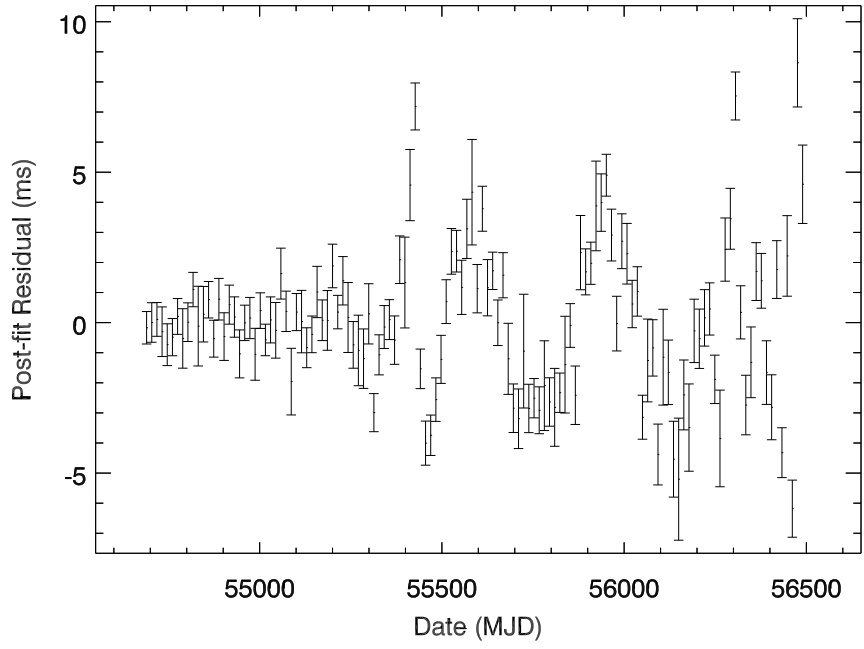}
\end{center}
\caption{Post-fit timing residuals for PSR J1907$+$0602.}
\label{fig:rms}
\end{figure}

\begin{figure}
\begin{center}
\includegraphics[width=5in]{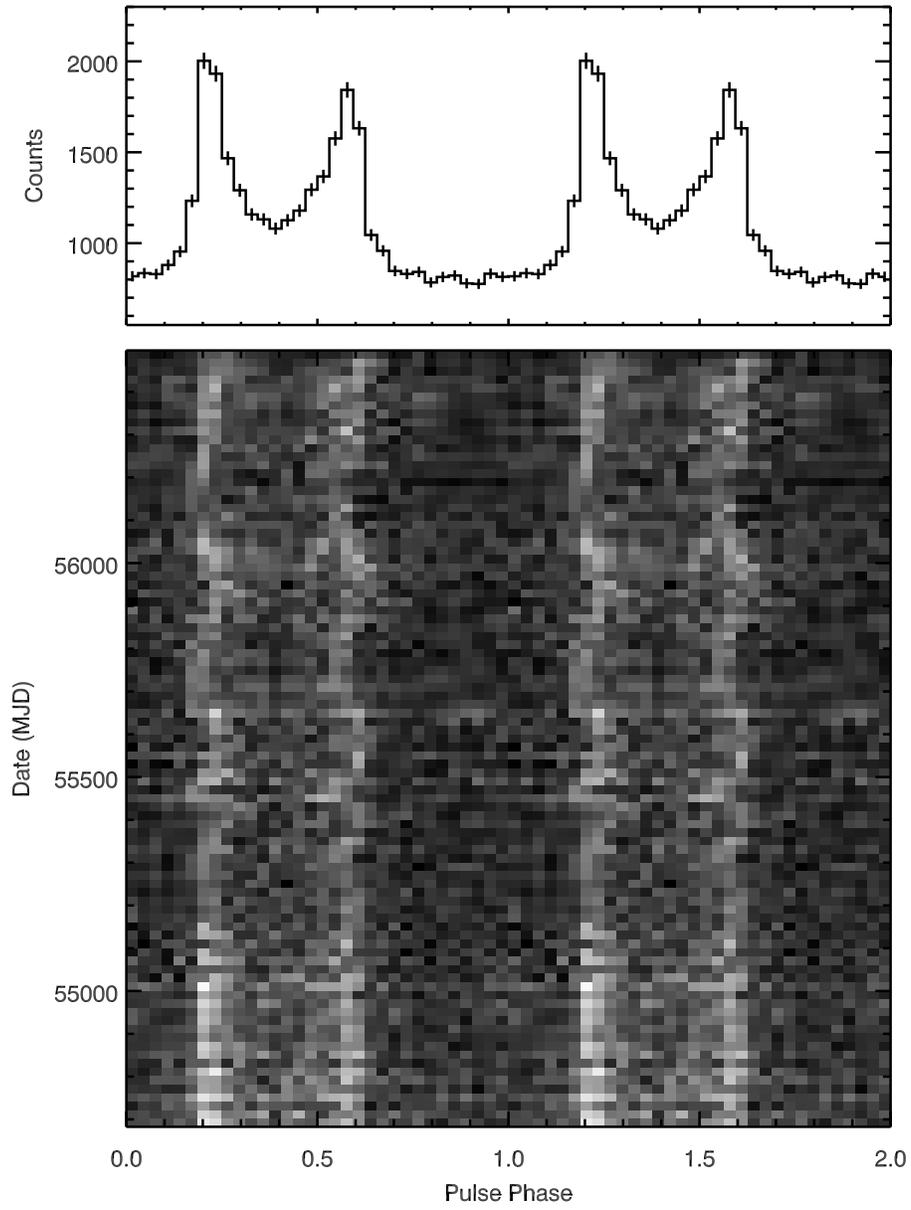}
\end{center}
\caption{Folded pulse profile and two-dimensional phaseogram in 32 phase bins obtained for PSR J1907$+$0602. For clarity, two rotations are shown on $X$-axis. The gray scale represents the number of photons in each bin.}
\label{fig:ftp}
\end{figure}

\begin{figure}
\begin{center}
\includegraphics[width=3in]{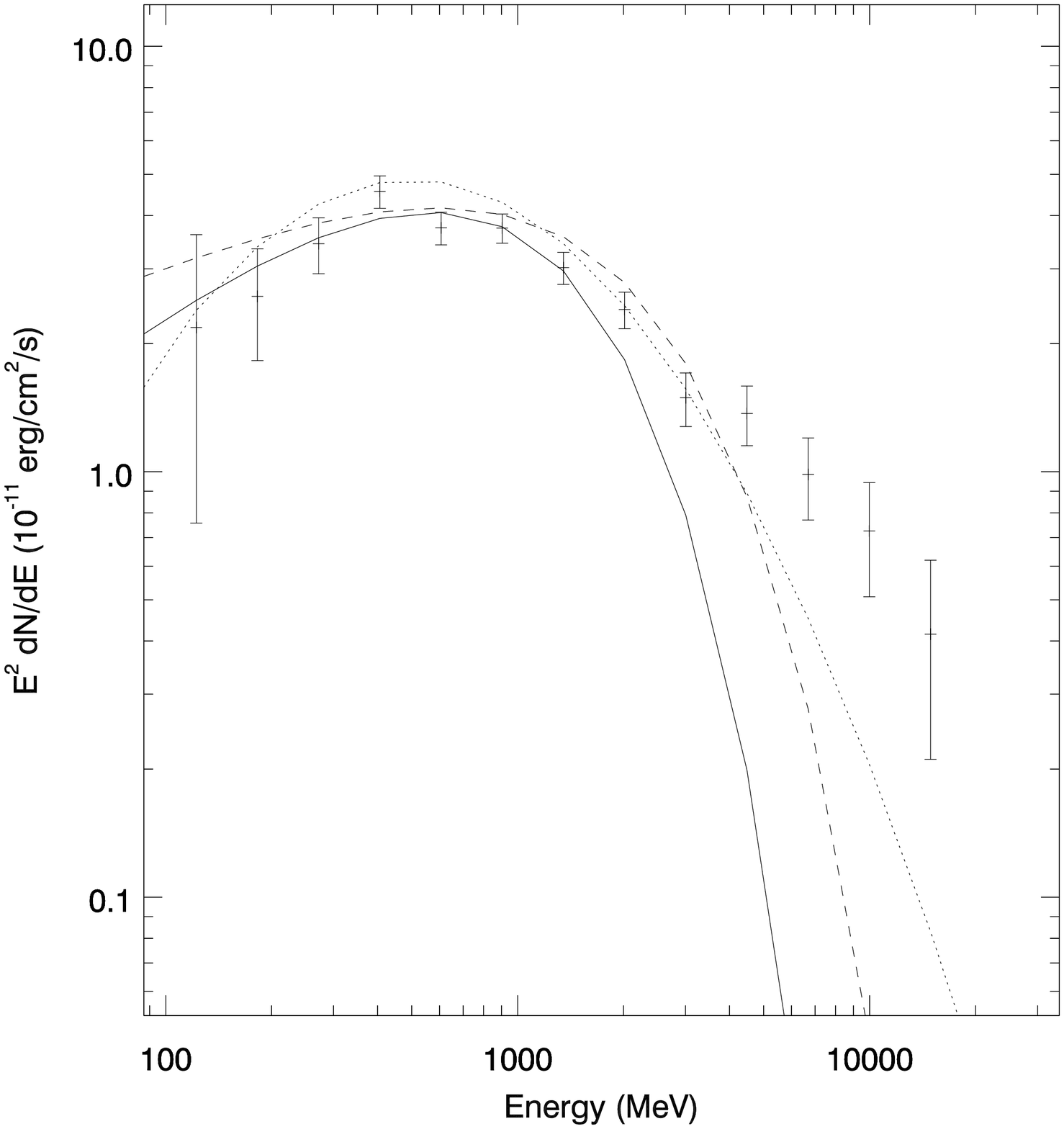}
\includegraphics[width=3in]{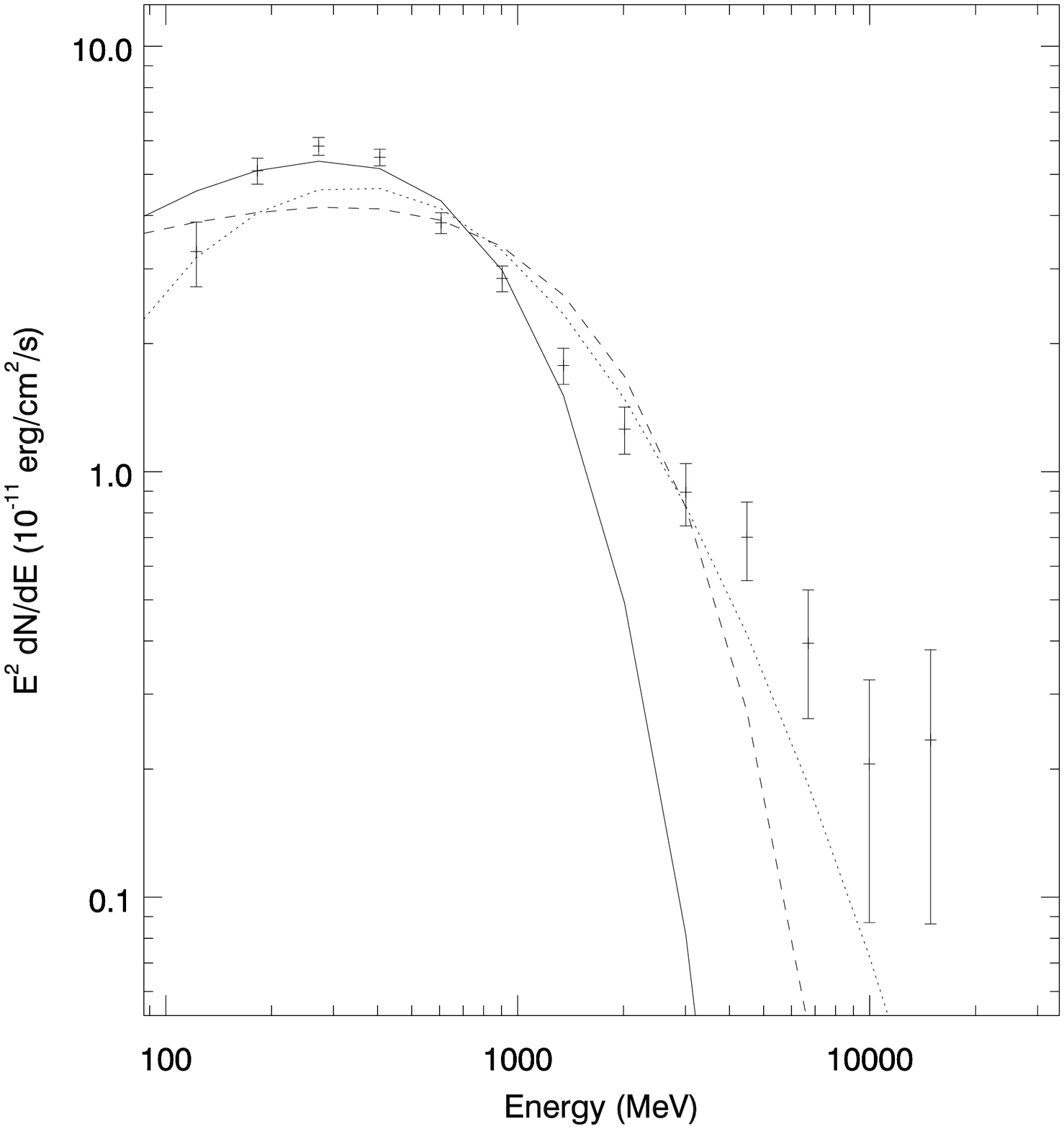}
\end{center}
\caption{$\gamma$-ray spectra of 2FGL J1906.5$+$0720 extracted from 
the full data ({\it left} panel) and offpulse phase interval data ({\it right}
panel). The exponentially cutoff power laws and the log parabolas 
obtained from maximum likelihood analysis (see Table~\ref{tab:likeli}) are 
displayed as dashed and dotted curves. The solid curves are the 
exponentially cutoff power laws obtained by fitting the data points 
below 2 GeV.}
\label{fig:spec}
\end{figure}

\begin{figure}
\begin{center}
\includegraphics[width=3in]{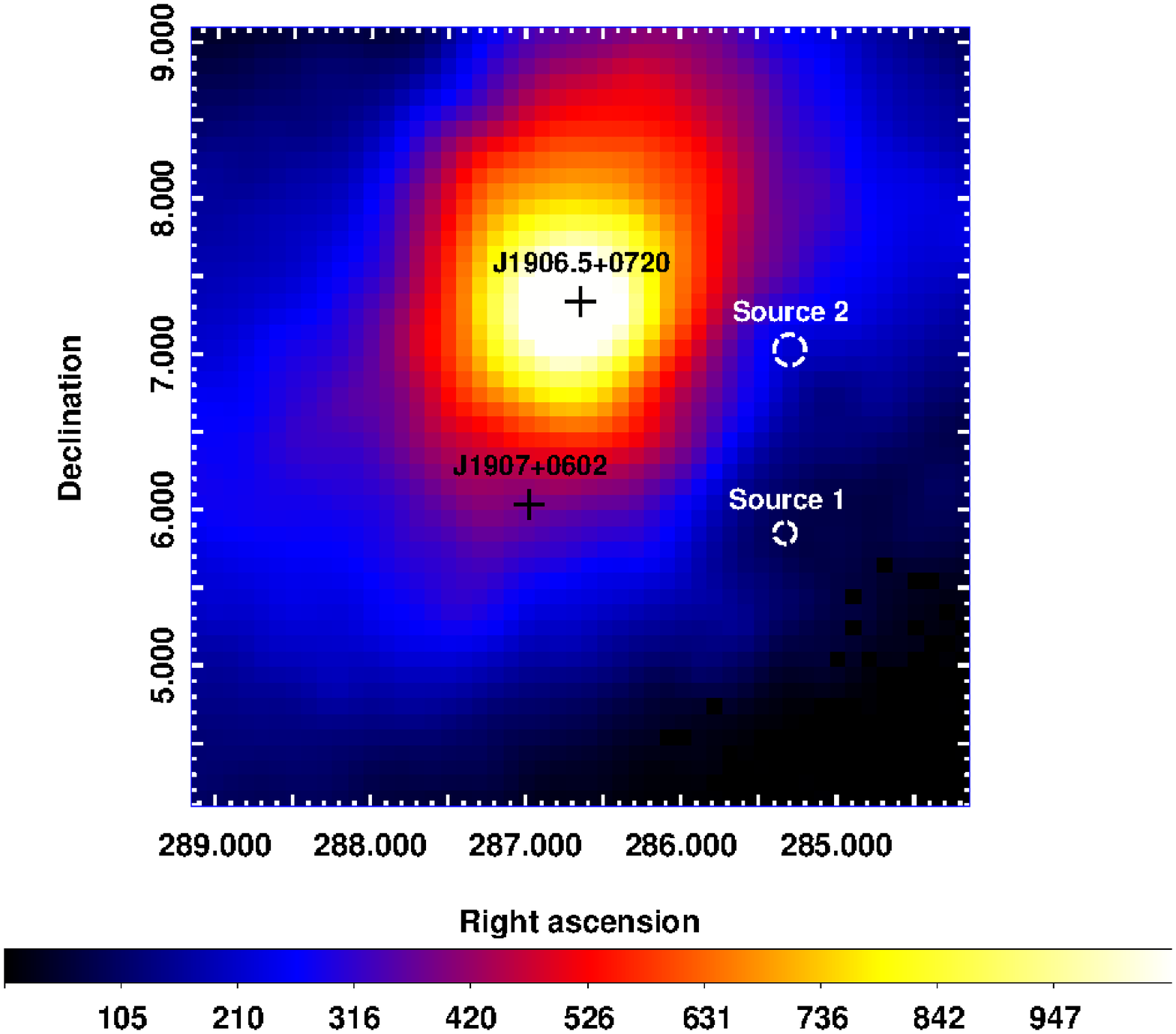}
\includegraphics[width=3in]{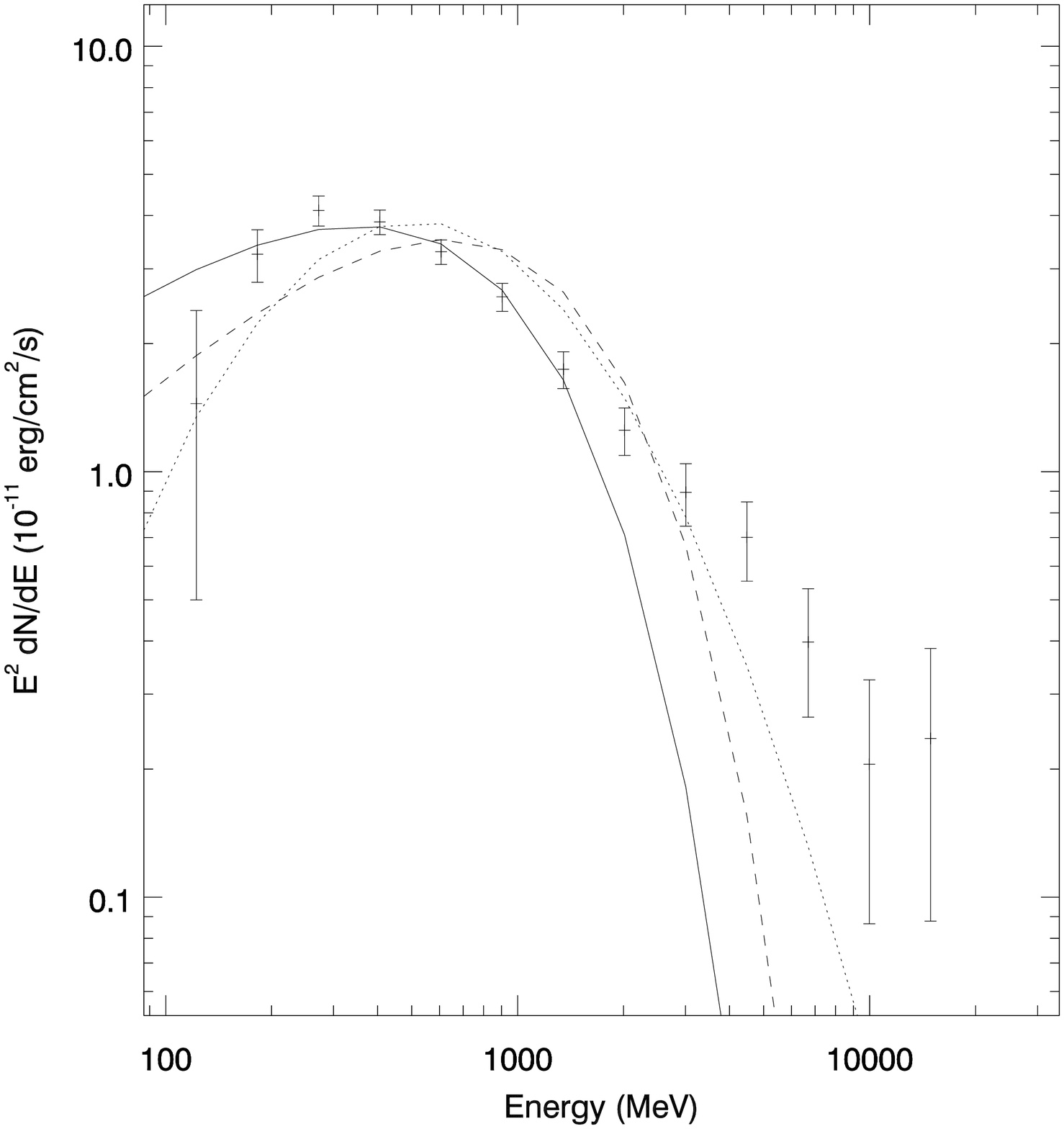}
\end{center}
\caption{{\it Left panel}: TS map (0.1--300 GeV) of 
$\mathrm{5^{o}\times5^{o}}$ region centered at 
R.A.= 286.647$\mathrm{^{o}}$, Decl.= 6.6$\mathrm{^{o}}$ (equinox J2000.0) 
extracted from the offpulse phase interval data of PSR J1907$+$0602 when the
two nearby sources (the positions are marked by circles) were removed. 
The image scale of the map is 0\fdg1 pixel$^{-1}$. {\it Right panel}: 
$\gamma$-ray spectrum of 2FGL J1906.5$+$0720 extracted from the offpulse 
phase interval data when the two nearby sources were removed. 
The dashed and dotted curves represent the exponentially cutoff 
power law and the log parabola, respectively, obtained from
maximum likelihood analysis (see Table~\ref{tab:likeli}). The solid curve
represents the exponentially cutoff power law obtained by fitting 
the data points below 2 GeV.}
\label{fig:notail}
\end{figure}

In the residual TS maps both from the full data and the offpulse phase interval 
data after removing all sources (Figure~\ref{fig:ts}), two
$\gamma$-ray emission excesses exist. They are located at 
R.A.=285\fdg326 and Decl.= 5\fdg855 (equinox J2000.0), with 
1$\sigma$ error circle of 
0\fdg07, and R.A.=285\fdg293 and Decl.= 7\fdg030 (equinox J2000.0), with 
1$\sigma$ error circle of 0\fdg1 (marked by circles in 
Figure~\ref{fig:ts}), which were obtained 
from running `gtfindsrc' in LAT science tools software 
package. In addition, there is also a tail-like structure in 
the southeast direction of 2FGL J1906.5$+$0720, which can be clearly 
seen in the TS map during offpulse intervals. In order to determine whether
this tail structure is associated with 2FGL J1906.5$+$0720 or caused by
the two nearby sources, we further performed maximum likelihood analysis
by including the two sources in the source model. 
The emission of the two putative sources were modeled by a simple power 
law.  We found that the tail structure was completely removed (see 
the left panel of Figure~\ref{fig:notail}), 
indicating that it is likely caused by the two nearby sources.

A $\gamma$-ray spectrum of 2FGL J1906.5$+$0720 was obtained again
for the offpulse phase interval data, with the two nearby sources 
considered. The three spectral models given in 
Section~\ref{subsec:likeli} were used. 
The results are given in Table~\ref{tab:likeli}. The spectral parameter
values are similar to those obtained above.
We also fit the obtained spectral data points below 2~GeV with 
an exponentially cutoff power law (cf. Section~\ref{subsec:sa}).
Nearly the same results were obtained (Table~\ref{tab:likeli}).

These analyses confirm the existence of a high-energy component in
the emission of 2FGL J1906.5$+$0720, which is likely not to be caused by
contamination from the nearby sources. By constructing TS maps with
photons greater than 2 or 5 GeV, we searched for extended emission
(e.g., a pulsar wind nebula) at the position of 2FGL J1906.5$+$0720.
However, the source profile was always consistent with being a point
source. There was no indication for the presence of an additional
source responsible for the high-energy component.

\subsection{Timing analysis of 2FGL J1906.5$+$0720}

Timing analysis was performed to the LAT data of 
2FGL J1906.5$+$0720 to search for $\gamma$-ray pulsation signals. 
We included events in the energy range from 50 MeV to 300 GeV within 1 degree 
centered at the position of 2FGL J1906.5$+$0720, which is
R.A.= 286\fdg647, Decl.= 7\fdg34256, equinox J2000.0 (the catalog position; 
\cite{nol2012}). The time period for the event selection was 300-day 
from 2012-09-26 20:53:17 to 2013-07-23 20:53:17 (UTC).
The time-differencing blind search technique described in \citet{atw2006} 
was applied.
The range of frequency derivative $\dot{\nu}$ over frequency $\nu$ we 
considered was $|\dot{\nu}/\nu |=0$--$1.3\times 10^{-11}$ $s^{-1}$, which is 
characteristic of pulsars such as the Crab pulsar. A 
step of $2.332\times 10^{-15}$ $s^{-1}$ was used in the search. 
The frequency range we considered was from 0.5 Hz to 32 Hz with 
a Fourier resolution of $1.90735\times 10^{-6}$ Hz. 
We did not include the parameter ranges characteristic of millisecond 
pulsars. The source 2FGL J1906.5$+$0720 is located in the Galactic 
plane and would be possibly a young pulsar such as PSR J1907$+$0602. 
No significant $\gamma$-ray pulsations from 2FGL J1906.5$+$0720 
were detected. 
We also applied the blind search to the \textit{Fermi}/LAT data 
of 2FGL J1906.5$+$0720 only 
during the offpulse phase intervals of PSR J1907$+$0602. 
No $\gamma$-ray pulsations except the spin frequency signal 
of PSR J1907$+$0602 were found.

In addition, we also searched for any long-period modulations from 
the source, the detection of which would be indicative of a binary system 
(see discussion in Section~\ref{sec:dis}).
We constructed power spectra during offpulse phase intervals 
of PSR J1907$+$0602 in the three energy bands of 0.2--1 GeV, 1--300 GeV, and 
5--300 GeV. Light curves of nearly five-year length in the three 
energy bands were extracted from performing \textit{Fermi}/LAT aperture 
photometry analysis. 
The aperture radius was 1 degree, and the time resolution of 
the light curves was 1000 seconds. The exposures were calculated assuming 
power law spectra with photon indices obtained by maximum likelihood 
analysis (Table~\ref{tab:likeli}), which were used to determine the flux 
in each time bin. No long-period modulations in the energy bands were found.

\section{Discussion}
\label{sec:dis}

By carrying out phase-connected timing analysis of the nearly 5-year
\fermi\ $\gamma$-ray data of PSR J1907$+$0602, we have 
obtained the timing parameters and updated the $\gamma$-ray ephemeris for
this pulsar.  
The obtained timing parameters are similar to those given in the second \fermi\ 
catalog of $\gamma$-ray pulsars \citep{lat2013}. However the glitch decay 
time constant is $\sim$99 days, larger than $\sim$33 days given
in the catalog. This difference is likely due to the longer time span of 
the data we analyzed (2-year data was analyzed in the second \fermi\ 
catalog of $\gamma$-ray pulsars; \cite{lat2013}) and the unstable 
timing parameters caused by the timing noise. PSR J1907$+$0602 
is quite young with a characteristic age of $\sim$19.5 kyr \citep{abd2010}.  
The post-fit rms timing residual was 2.1 ms, resulting from
our timing analysis (Table~\ref{tab:timing}). 

We performed different analyses of the \fermi/LAT data for the unassociated 
source 2FGL J1906.5$+$0720. Through likelihood analysis with different spectral
models, we confirmed that a curved 
spectrum with a low-energy break or cutoff at $\sim$1 GeV is clearly preferred 
to a simple power law. The significances of 
the curvature ($\sim\sqrt{TS}\sigma$) are approximately 14--16 $\sigma$ 
and 16--18 $\sigma$ for the full data and the offpulse phase interval data,
respectively. This feature is characteristic of $\gamma$-ray pulsars 
detected by \textit{Fermi}. 
On the basis of the \fermi\ second pulsar catalog, 
young $\gamma$-ray pulsars have 0.6 $<\Gamma<$2 and 
0.4 GeV $< E_{cutoff} <$ 5.9 GeV, and 
millisecond $\gamma$-ray pulsars have 0.4 $<\Gamma <$ 2 and 
1.1 GeV $< E_{cutoff} <$ 5.4 GeV (\cite{lat2013}). 
If 2FGL J1906.5$+$0720 is a pulsar, its Galactic location and
spectral feature suggest that it is probably a young pulsar
(see, e.g., \cite{lat2013}). 
It should be noted that a log parabola, which better fits the spectra 
of 2FGL J1906.5$+$0720, is usually used to model the spectra of $\gamma$-ray 
binaries \citep{nol2012}. However, considering the non-detection of any long-period modulations and 
the low variability of 2FGL J1906.5$+$0720, a $\gamma$-ray binary is not
likely the case for the source.

From our spectral analysis, a high-energy component was found to
exist at $\geq$4 GeV in the emission of 2FGL J1906.5$+$0720. 
Considering it as a young pulsar,
the component likely originates from its pulsar wind nebula 
(PWN; e.g., \cite{gs06}).
A pulsar wind generates a termination shock by the interaction of
high-energy particles contained in it with the ambient medium, at which 
particles are re-distributed and can radiate
ultra-relativistic emission. In the $\gamma$-ray energy range,
the \fermi\ second source catalog used 
69 known PWNe for the automatic source association, and found that 
nearly all of them (except three) are associated with young 
pulsars \citep{nol2012}. However since $\gamma$-ray emission from a pulsar
often dominates over that from its PWN, the number of PWNe that have been
confirmedly detected by \fermi\ is limited \citep{ack2011}.
For 2FGL J1906.5$+$0720, our spatial distribution analysis has confirmed
the existence of the high-energy component in its spectrum, but the putative
PWN would be too small or too faint to be resolved by \fermi. Further X-ray
imaging of the source field is needed in order to detect the PWN and thus
help verify the pulsar nature for 2FGL J1906.5$+$0720.

We have not been able to find any pulsed emission signals from
the \fermi\ data of 2FGL J1906.5$+$0720, which is required to
verify the source's pulsar nature. We note that the LAT blind 
search sensitivity
depends on many parameters, such as the accurate position of the source,
the source region used for pulsation search, contamination from background
diffuse emission and from nearby sources (given that our target is located
at the Galactic plane with several identifiable sources nearby). 
Using the sensitivity estimation method for the blind searches provided
by \citet{dor2011}, the pulsed fraction of 2FGL J1906.5$+$0720 should be 
$\gtrsim$0.57 for a detection probability of $>$68\% (the 1-year detection 
significance is $\sim$20$\sigma$ for the source). \citet{dor2011} also 
extracted an all-sky detectability flux map to describe the minimum 
0.3 -- 20 GeV photon flux required for the detection of pulsars with pulsed
fractions. In the inner Galactic plane the detectability flux should be 
higher than $\sim$10$^{-7}$ ph cm$^{-2}$ s$^{-1}$. 
The 0.3 -- 20 GeV photon flux we obtained for 2FGL J1906.5$+$0720 is 
$<$10$^{-7.1}$ ph cm$^{-2}$ s$^{-1}$ (derived from spectral parameters 
listed in Table~\ref{tab:likeli}), suggesting the difficulty of detecting 
pulsed emission from the source through blind searches. 
Considering the radio pulsations from the source have been searched 
several times but with no detection \citep{ray2012}, in order
to verify its pulsar nature, X-ray observations are needed. 

\bigskip

We thank the referee for valuable suggestions. This research was supported by Shanghai Natural Science Foundation for
Youth (13ZR1464400), National Natural Science Foundation of China (11373055), and the Strategic Priority Research Program ``The Emergence of Cosmological Structures" of
the Chinese Academy of Sciences (Grant No. XDB09000000). ZW is a Research Fellow of the One-Hundred-Talents project of Chinese Academy of Sciences.

\clearpage
\begin{table}
\caption{The first 10 candidate pulsars ranked by Signif\_Curve}\label{tab:candi}
\begin{center}
\begin{tabular}{lccccccc}
\hline
\hline
Source & Signif\_Curve ($\sigma$) & $Gb$ (\textdegree) & Variability\_Index & Signif\_Avg ($\sigma$) \\
\hline
2FGL J1704.9$-$4618 & 10.0 & $-$3.111 & 21.3 & 9.3 \\
2FGL J1906.5$+$0720 & 9.8 & $-$0.002 & 30.9 & 24.0 \\
2FGL J1819.3$-$1523 & 9.2 & $-$0.072 & 30.0 & 19.3 \\
2FGL J1847.2$-$0236 & 8.5 & $-$0.257 & 31.3 & 13.8 \\
2FGL J1856.2$+$0450c & 8.4 & 1.139 & 18.8 & 12.3 \\
2FGL J1619.0$-$4650 & 8.3 & 2.457 & 22.2 & 10.6 \\
2FGL J2033.6$+$3927 & 8.3 & $-$0.382 & 33.8 & 13.0 \\
2FGL J1045.0$-$5941 & 8.3 & $-$0.639 & 21.5 & 36.1 \\
2FGL J0858.3$-$4333 & 8.1 & 1.428 & 16.5 & 14.1 \\
2FGL J1739.6$-$2726 & 8.1 & 1.906 & 27.6 & 15.2 \\
\hline
\end{tabular}
\end{center}
\end{table}

\clearpage
\begin{table}
\caption{$\gamma$-ray ephemeris for PSR J1907$+$0602.}\label{tab:timing}
\begin{center}
\begin{threeparttable}
\begin{tabular}{lc}
\hline
\hline
Parameter & Value\tnote{*} \\
\hline
R.A., $\alpha$ (J2000.0) & 19:07:54.7343205 \\
Decl., $\delta$ (J2000.0) & 06:02:16.97850 \\
Pulse frequency (s$^{-1}$) & 9.3776609432(10) \\
Frequency first derivative (s$^{-2}$) & $-$7.62737(7) $\times$ 10$^{-12}$ \\
Frequency second derivative (s$^{-3}$) & 1.95(2) $\times$ 10$^{-22}$ \\
Epoch of frequency (MJD) & 55422.275976 \\
Dispersion measure (cm$^{-3}$ pc) & 82.1 \\
1st glitch epoch (MJD) & 55422.155 \\                          
1st glitch permanent frequency increment (s$^{-1}$) & 4.3466(3) $\times$ 10$^{-5}$ \\ 
1st glitch frequency deriv increment (s$^{-2}$) & $-$7.72(2) $\times$ 10$^{-14}$ \\
1st glitch frequency increment (s$^{-1}$) & 2.10(5) $\times$ 10$^{-7}$ \\   
1st glitch decay time (Days) & 99(4) \\
rms timing residual (ms) & 2.1 \\
Time system & TDB \\
E$_{min}$ & 50 MeV \\
Valid range (MJD) & 54683--56497 \\
\hline                              
\end{tabular}
\begin{tablenotes}
\footnotesize
\item[*] Parameters with no uncertainty reported are fixed to the values given in the second \textit{Fermi} catalog of $\gamma$-ray pulsar \citep{lat2013} except the glitch epoch.
\end{tablenotes}
\end{threeparttable}
\end{center}
\end{table}

\clearpage
\begin{table}
\caption{Maximum binned likelihood results for 2FGL J1906.5$+$0720}\label{tab:likeli}
\begin{center}
\begin{threeparttable}
\begin{tabular}{llcccccc}
\hline
\hline
Spectral model & Parameters & Full data & Offpulse phase interval data & Offpulse phase interval data\\
 & & & & (`tail' removed) \\
\hline
PowerLaw & $\Gamma$ & 2.31 $\pm$ 0.02 & 2.42 $\pm$ 0.02 & 2.31 $\pm$ 0.02 \\
 & $G_{\gamma}$ (10$^{-11}$ erg cm$^{-2}$ s$^{-1}$) & 15 $\pm$ 0.5 & 13 $\pm$ 0.4 & 11 $\pm$ 0.4 \\
 & TS$_{PL}$ & 1101 & 2437 & 1595 \\
\hline
LogParabola & $\alpha$ & 2.52 $\pm$ 0.05 & 2.82 $\pm$ 0.05 & 2.73 $\pm$ 0.07 \\
 & $\beta$ & 0.35 $\pm$ 0.03 & 0.37 $\pm$ 0.03 & 0.51 $\pm$ 0.04 \\
 & $E_{b}$\tnote{*} (GeV) & 1 & 1 & 1 \\
 & $G_{\gamma}$ (10$^{-11}$ erg cm$^{-2}$ s$^{-1}$) & 13 $\pm$ 0.5 & 12 $\pm$ 0.4 & 9 $\pm$ 0.3 \\
 & TS$_{LP}$ & 1388 & 2795 & 1966 \\
\hline
PLSuperExpCutoff & $\Gamma$ & 1.7 $\pm$ 0.2 & 1.7 $\pm$ 0.1 & 1.2 $\pm$ 0.1 \\
 & $E_{cut}$ (GeV) & 1.7 $\pm$ 0.8 & 1.2 $\pm$ 0.2 & 0.8 $\pm$ 0.1 \\
 & $G_{\gamma}$ (10$^{-11}$ erg cm$^{-2}$ s$^{-1}$) & 13 $\pm$ 5 & 12 $\pm$ 2 & 9 $\pm$ 1 \\
 & TS$_{PL+cutoff}$ & 1297 & 2694 & 1853 \\
\hline
PLSuperExpCutoff & $\Gamma$ & 1.4 $\pm$ 0.3 & 1.4 $\pm$ 0.1 & 1.4 $\pm$ 0.2 \\
obtained by fitting & $E_{cut}$ (GeV) & 1.0 $\pm$ 0.3 & 0.5 $\pm$ 0.1 & 0.6 $\pm$ 0.1 \\
\hline
\end{tabular}
\begin{tablenotes}
\footnotesize
\item[*] The break energies are fixed at 1~GeV.
\end{tablenotes}
\end{threeparttable}
\end{center}
\end{table}

\clearpage

\clearpage

\clearpage

\clearpage

\clearpage

\end{document}